\documentclass[11pt,twoside]{article}

\usepackage{asp2006}
\usepackage{epsf}
\usepackage{lscape}
\usepackage{graphicx}

\begin{document}
\title{Spontaneous and Stimulated Star Formation in Galaxies: Ultraviolet
Limits on Star Formation Thresholds and Optical Constraints on
Lambda-CDM Cosmological Simulations of Galaxy Formation.
}   
\author{Barry F. Madore}
\affil{The Observatories \\
Carnegie Institution of Washington \\
and \\
NASA/IPAC Extragalactic Database (NED) \\
California Institute of Technology \\
Pasadena CA}    

\author{Samuel Boissier}
\affil{Labratoire d'Astrophysique de Marseille\\
Marseille, France}    

\author{Armando Gil de Paz}
\affil{Dept. de Astrofisica \\
Universidad Complutense de Madrid \\
Madrid, E-28040 Spain}    

\author{Erica Nelson \& Kristen Petrillo }   
\affil{The Observatories \\
Carnegie Institution of Washington \\
and \\
Pomona College \\
Claremont, CA}    

\begin{abstract} 

We present recent results from several on-going studies: The first
addresses the question of gas-density thresholds for star formation,
as probed by the outer disks of normal nearby galaxies. The second
concerns the observational evidence for the existence of gravitating
non-luminous (GNL) galaxies, as predicted by most recent simulations
of galaxy formation in Lambda-CDM cosmologies. We find that (1) If
star formation is traced by far-ultraviolet light, then there is no
evidence for a threshold to star formation at any gas density so far
probed, and (2) there is no evidence for GNL galaxies gravitationally
interacting with known optical systems based on the observations (a)
that there are no ring galaxies without plausible optically visible
intruders, (b) all peculiar galaxies in the Arp Atlas that are bodily
distorted have nearby plausibly interacting companions, and (c) there
are no convincingly distorted/peculiar galaxies within Karachentsev's
sample of more than 1,000 apparently/optically isolated galaxies.

\end{abstract}


\section{Introduction}   

Galaxies continue to surprise us, both in their observed structure and
by their inferred evolution. Some of these surprises come from moving
to new wavelengths and re-observing old friends; while other surprises
come staying at familiar wavelengths but looking at the galaxies from
a somewhat new perspective or in a slightly revised context.

With the launch of the GALEX satellite into low-Earth orbit on April
28, 2003 it has become possible to image a large number and a wide
range of different types of galaxies at largely unexplored
wavelengths: in the near and far ultraviolet. Given the high
sensitivity of the detectors and the very low sky background
(especially in the far ultraviolet channel) it is possible to see
features in the ultraviolet out to surface brightness levels
unequalled by ground-based optical observations. Moreover, the wide
(one-degree) field of view of GALEX also maximizes the possibility of
serendipitous discovers, of which there have been many.

\section{Radially Extended Star Formation}

All of the above aspects of GALEX contributed to the ``discovery'' of
{\it extended ultraviolet} (XUV) disks around a number of nearby
spirals.  Prime examples, found early in the mission are M83 (Thilker
{\it et al.} 2005) and NGC~4625 (Gil de Paz {\it et al.} 2005). A
comparison of the optical image of the compact, one-armed spiral NGC~4625
with its significantly more extended, and multi-armed UV counterpart
image is shown in Figure 1.

\begin{figure}[!ht]
\begin{center}
\includegraphics[scale = 0.6] {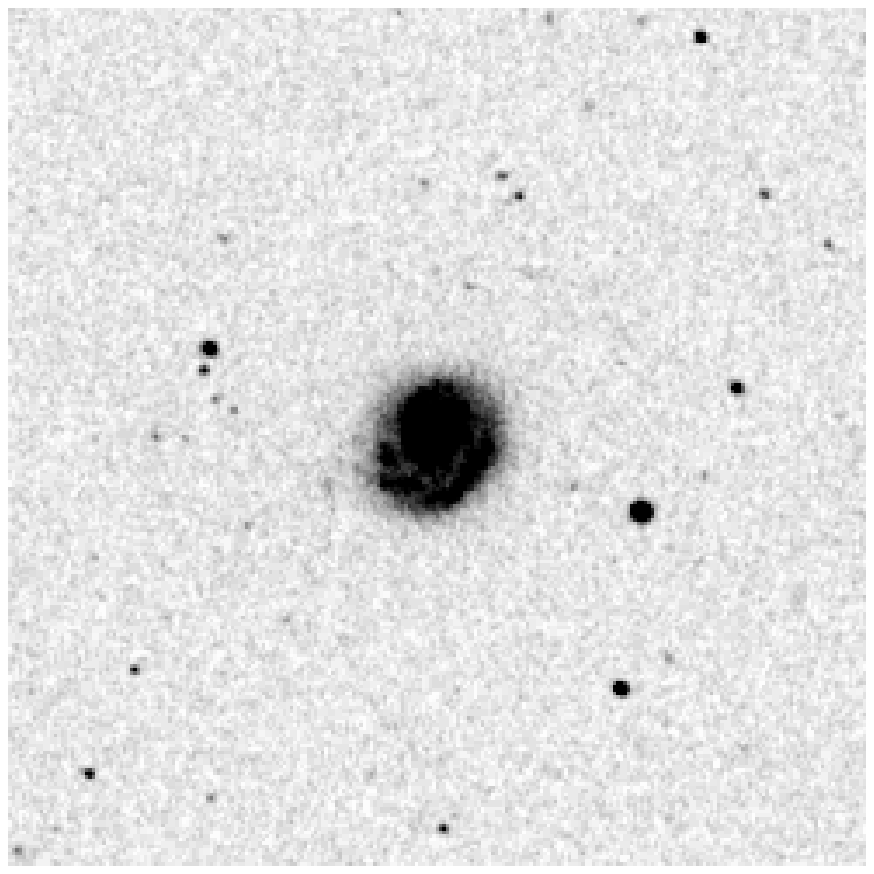}
\includegraphics[scale = 1.1] {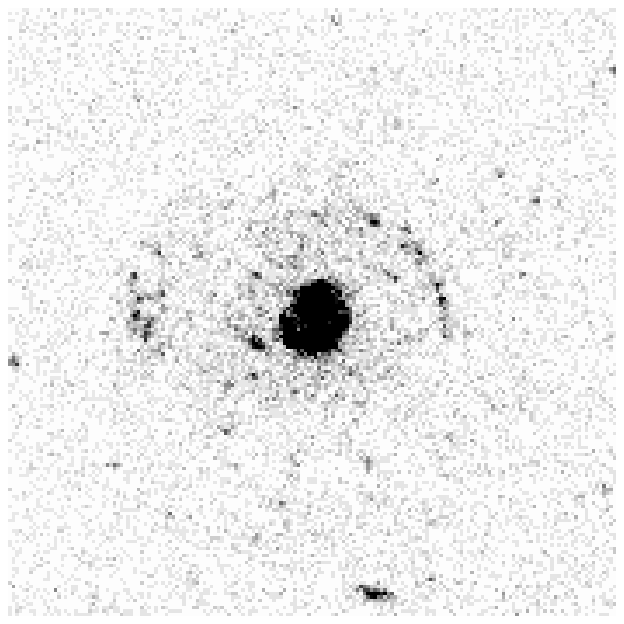}
\end{center}
\caption{NGC 4625. Right panel shows the optical image of the galaxy,
while the left panel shows the NUV (GALEX) of this same galaxy where
the extended UV disk of star formation is clearly
highlighted}\label{fig1}
\end{figure}

\subsection{Pre-Discoveries and Thresholds}

While these features are striking, they should not have come as a
surprise (althogh they did) as they are not unheralded. For instance,
based UV imaging data obtained from a balloon-borne experiment (SCAP
2000) Donas et al. (1981) asked the question ``How Far Does M101
Extend?'' Clearly they were seeing extended UV features in the outer
disk of a well known nearby galaxy.  And later Ferguson et al. (1998),
announced ``The Discovery of Recent Star Formation in the Extreme
Outer Regions of Disk Galaxies'' based on deep H$\alpha$ surveys of
three nearby late-type galaxies: NGC~0628, NGC~1058 and NGC~6946. And
all of this might have been seen to be a natural consequence of the
discovery of extended HI disks around many galaxies (e.g., NGC~2915,
Muere et al. 1996) except perhaps for the interesting interpretive
paper by Martin \& Kennicutt (2001) which seemed to put an end to star
formation in the outer disks of galaxies by titling their paper ``Star
Formation Thresholds in Galactic Disks.'' With hindsight, which we all
now possess, it might have been more robust to have added ``as Traced
by HII Regions'' to the title.

\subsection{Ultraviolet Data}

Recently, using GALEX ultraviolet imaging data, Boissier et al. (2007)
have re-examined the correlation of gas density with star formation
({\it as traced by FUV/NUV light}) and have come to very different
conclusions regarding the existence of any threshold to star formation
at low gas densities. As can be seen in Figure 2 the left panel shows
the run of star formation rates as a function of total gas density for
a combined sample of 43 galaxies. The star formation rate is based on
extinction-corrected UV surface brightness obtained by the GALEX
satellite and the total gas density combines neutral and molecular
hydrogen contributions.  The right panel shows a selection of
individual galaxies where the run of UV and the H$\alpha$ star
formation rates with surface density of gas are intercompared on an
individual basis. Clearly there is a difference. No truncation and no
threshold is apparent in the UV data. The surface density of star
formation as traced by hot, blue stars (which may or may not) include
O stars (which power the brightest HII regions) is continuous with the
gas surface density, to the limits of both surveys.

The hard cut-off in H$\alpha$ has been challenged, of course, by
Ferguson et al. (1988), but it may also be that other factors are in
play.  In the outer parts of galaxies we may be seeing small number
statistics force the mass function of the molecular clouds down to a
mass level that while they can support star formation they are not
individually large enough to produce even a single O star capable of
ionizing the surrounding medium. The plane thickness may have grown so
much, or the intercloud separation may be so large that the star
formation regions are density-bounded and that large fraction of the
UV radiation is leaking out before it can produce a detectable HII
region or that the radiation that is intercepted is only reradiated at
a very low emission measure and perhaps and widely distributed. Very
compact HII regions have been found in the XUV disk of M83 and they
are consistent with single-star ionization (Gil de Paz et
al. 2005). Studies are underway to probe deeper and to lower surface
brightness levels in search of any stray radiation.

\begin{figure}[!ht]
\includegraphics[scale = 0.360] {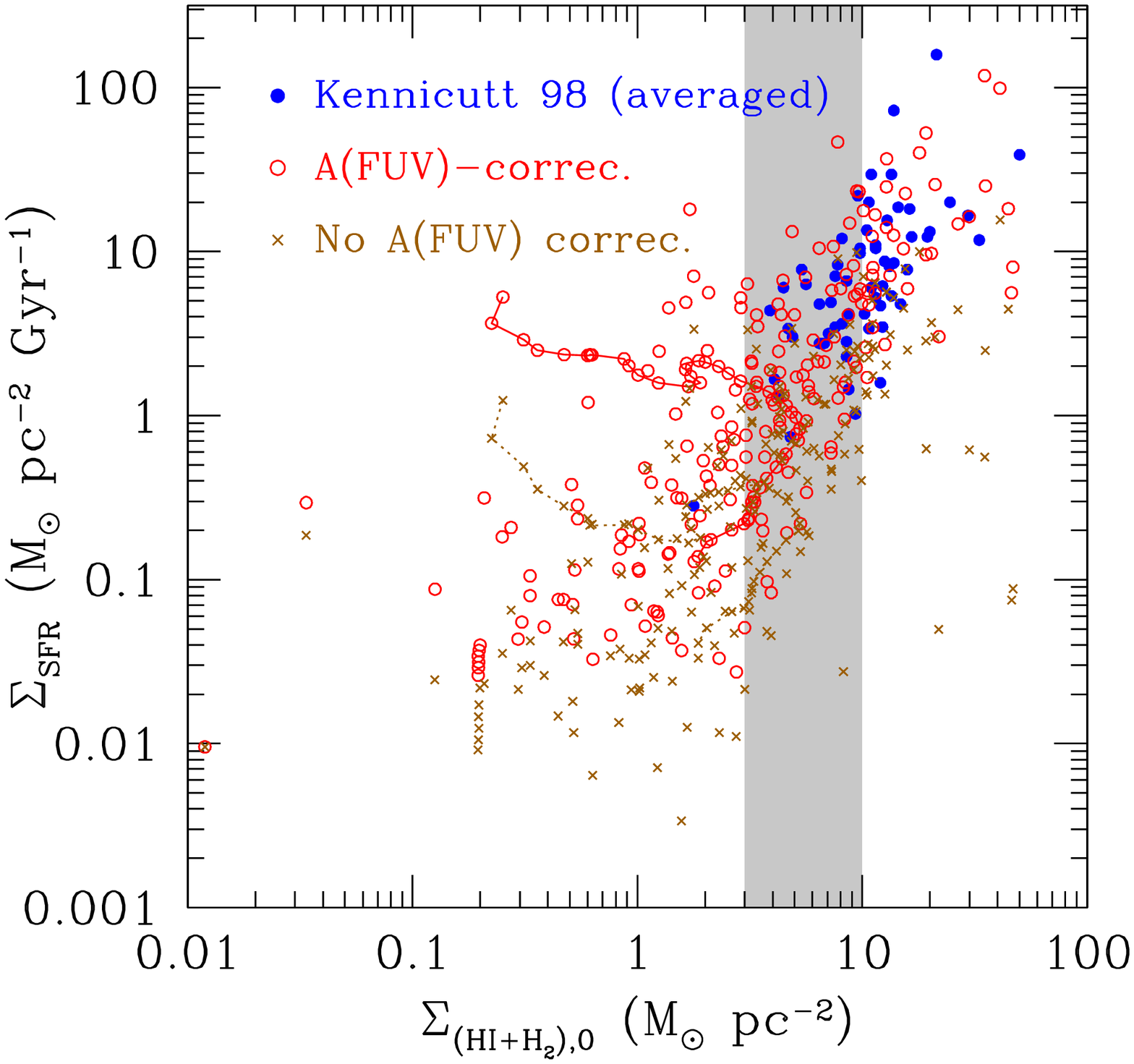}
\includegraphics[scale = 0.340, bb = 45 110 65 145] {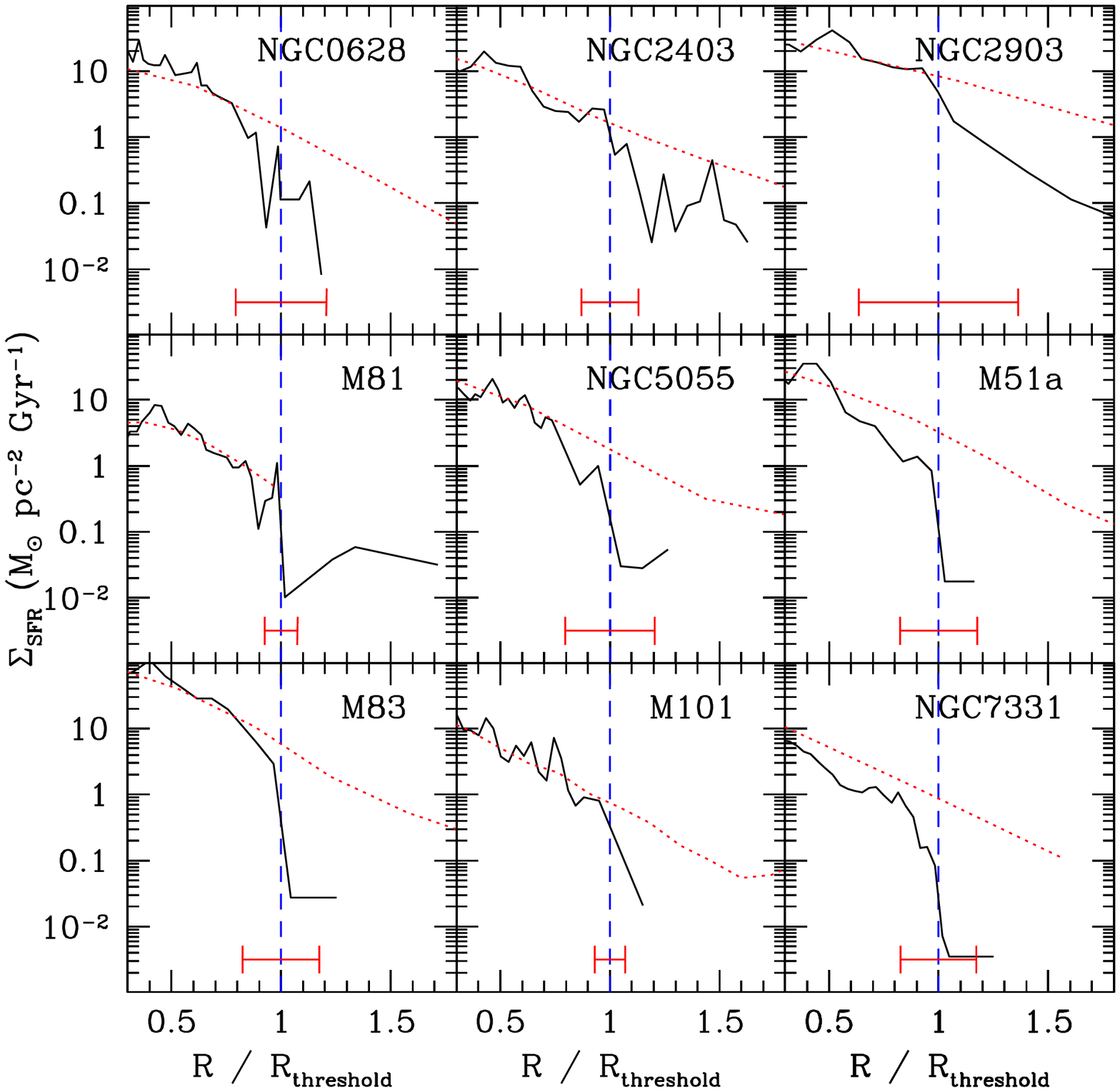}
\caption{The left panel shows the radial drop-off of star formation
rates as a function of gas density for ** galaxies as reproduced from
Boissier et al. (2007). The vertical grey zone shows the gas-density
where the star-formation threshold of Martin \& Kennicutt (2001)is
expected. For the UV data no threshold is observed. The right panels
shows an intercomparison of UV star formation rates (dotted lines) and
H$\alpha$ star formation rates (solid lines) as a function of radius
(scaled to the Martin-Kennicutt threshold.) Again, the UV star
formation shows no sign of any thresholding at any gas density
plotted.}\label{fig3}
\end{figure}

\section{Gravitating Non-Luminous (GNL) Galaxies}

Most contemporary models of structure formation within the framework
of a $Lambda$-CDM cosmology predict a steep power-law increase of
lower mass galaxies fainter than $L^*$ (e.g., Kravtsov, Gnedin \&
Klypin 2004). Most of the satellites are deemed ``missing'' because
they not seen in optical surveys. This has led to the speculation that
they are currently somehow devoid of baryons, and thus invisible by
construction. Invisible but not undetectable. The presence of {\it
Gravitating Non-luminous} (GNL) galaxies can in principle (and in
practice) be inferred (and seen) by their gravitational interactions
with other nearby (visible) galaxies. Below we explore a few such
tests for their presence ... or absence.

\subsection{Ring Galaxies}

Twenty years ago Arp \& Madore (1987) published a catalog and
photographic atlas of peculiar galaxies based on a visual inspection
of over 94,000 optical images of southern-hemisphere galaxies. More
recently Madore, Nelson \& Petrillo (2007 in prep) extracted from that
a pure sample of about 100 ring galaxies (see Figure 3 for a
sampling). According to models by Theys \& Spiegel (1976) and by Lynds
\& Toomre (1976) they can be explained by galaxy-galaxy interactions,
fine-tuned to be head-on collisions between a disk galaxy and a
lower-mass intruder. The ring galaxies culled from the Arp-Madore
Catalogue were listed by those original discoverers because of their
ring morphology not because they did or did not have
companions. However, it is gratifyingly strong confirmation of the
theory that virtually all of the rings also have adjacent (line of
sight) companions (many of which already have redshifts confirming
their physical as well as apparent association with the rings.) These
companions are plausible colliders, based on their apparent proximity
and being only a diameter or two from the ring itself.
\begin{figure}[!ht]
\begin{center}
\includegraphics[scale = 0.55] {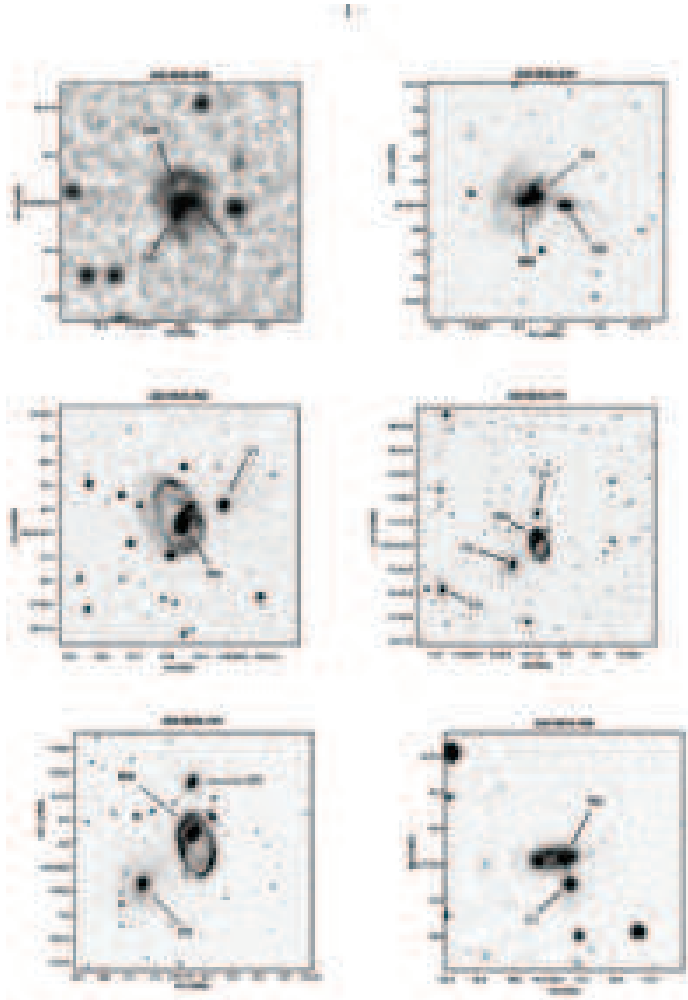}
\end{center}
\caption{Examples of ring galaxies and their adjacent companions from
the soon to be published atlas of Madore, Nelson \& Petrillo
(2007)}\label{fig2}
\end{figure}

While this is all good news for the theory of ring formation in
specific, it is not so good news for the theory of galaxy formation in
general. Without a single convincing example of an isolated ring, one
obvious conclusion is that GNL galaxies do not exist in the numbers
predicted by theory. While it may be counter-argued that ring galaxies
require such peculiar circumstances for their formation (mass ratios,
orbital parameters and galaxy types, etc.) and that they can only
arise from a collision involving concentrated {\it baryonic} satellite
intruders for their formation, the same cannot be said for
collision-induced peculiarities in general. The CDM simulations
predict the observed number of optical galaxy only at one mass. Above
and below this there is a divergence at all masses bewteen theory and
observation, reach more than a factor of 10 descrepancy at the fainest
end. Wehether is is high-mass or low-mass companions that are being
looked for in GNL-galaxy interactions they are predicted to be there
at the level of factors more than their optical counterparts rather
than occasionally occurring at very low levels of incidence, as
appears to be the case.

\subsection{Arp Peculiar Galaxies}

Turning back to the time at which the original {\it Atlas of Peculiar
Galaxies} was published (Arp 1966) it is fair to say that the topic of
peculiar galaxies was still in its formative stages and that the
sample illustrated was not premised upon their being or not being
nearby galaxies to qualify them to be include in the Atlas. Indeed by
moder standards many of the galaxies in the {\it Atlas} are not now
considered to be particularly peculiar at all ({\it e.g.,}
low-surface-brightness galaxies [ARP 001-004], dwarf irregular
galaxies [ARP 005-006], and certain alignments in small groups and
clusters [ARP 311-332]). Rare does not necessarily mean peculiar, but
the {\it Atlas of Peculiar Galaxies} did include many rare types of
objects. Our point here is however that of the 338 objects that are
included in the {\it Atlas} because the galaxy in question is bodily
distorted, is considerably asymmetric or has extended tails, the vast
majority of those have very nearby companions that can easily be
implicated as the source of the interaction and distortion. It is the
absence of {\it isolated} peculiar galaxies that is in itself
noteworthy. Clearly the vast majority of peculiarities seen as bodily
deformations of the galaxy in question can be explained by
interactions with nearby optical companions. And this simple fact
leaves no room (and no evidence for) bodily deformed galaxies
resulting from their interaction with gravitating non-luminous (GNL)
galaxies (i.e., pure dark-matter halos).
\begin{figure}[!ht]
\begin{center}
\includegraphics[scale = 0.5] {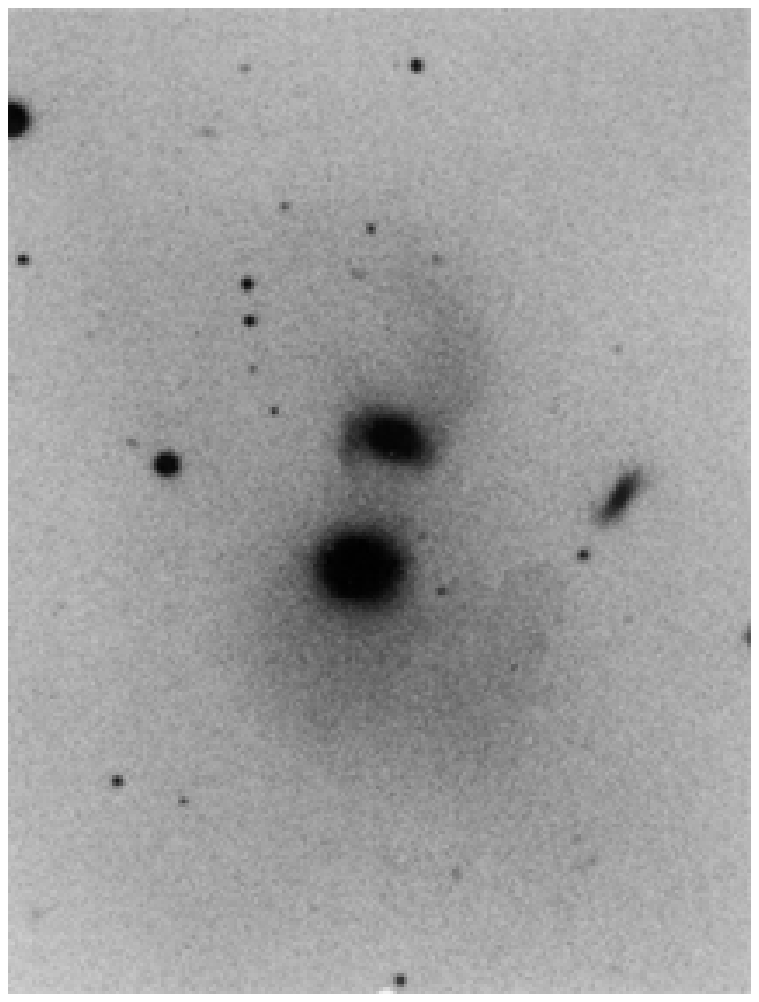}
\includegraphics[scale = 0.5] {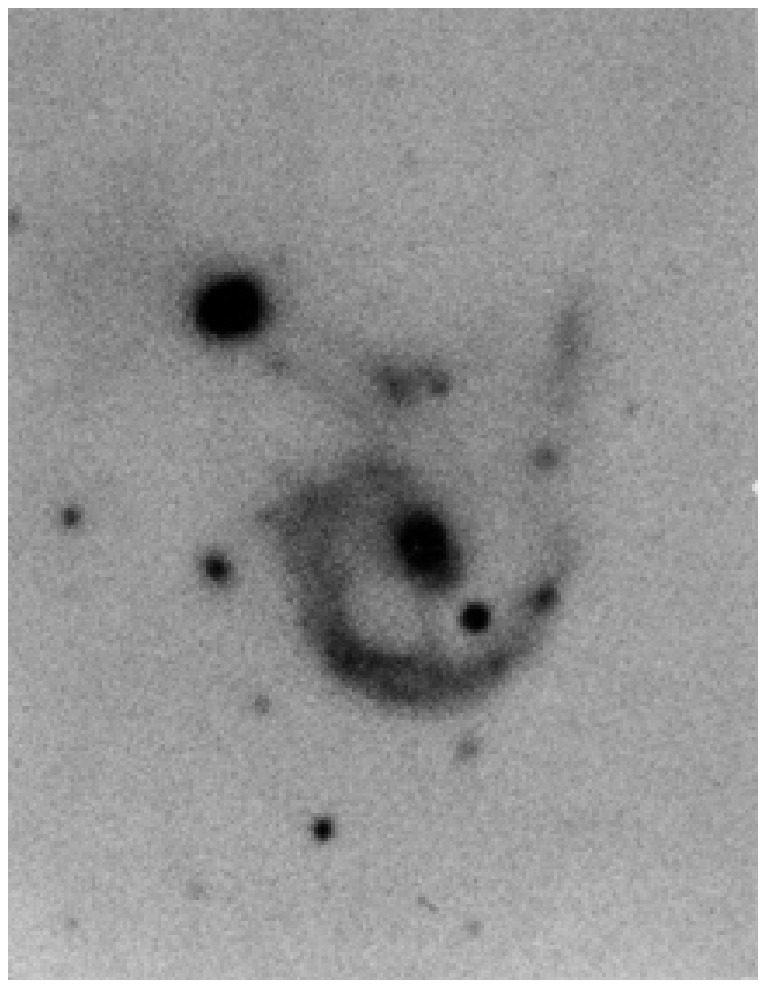}
\includegraphics[scale = 0.5] {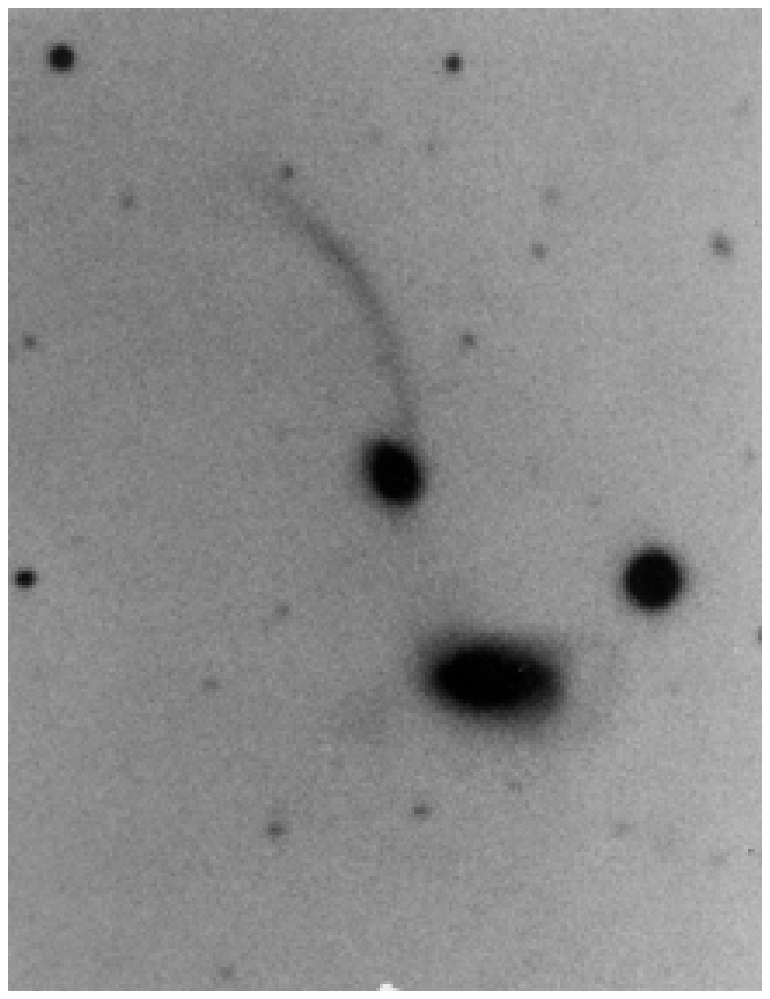}
\end{center}
\caption{Arp~172 (left), Arp~107 (center) and Arp ~173 (right) typify
the types of bodily distorted galaxies in the Arp Atlas of Peculiar
Galaxies that almost without exception have obvious interactions
on-going between two optically visible galaxies}\label{fig1}
\end{figure}

\subsection{Karachentsev Isolated  Galaxies}

There is an alternative path to follow here. Karachentsev (1988)
published a list of 1,000 galaxies that are optically isolated from
other comparable-sized (optically visible) galaxies. This, of course,
is not to say that apparently isolated cannot and do not have GNL
galaxies of comparable (or even larger) sizes orbiting and interacting
with them.  However this catalog would suggest that this is not
the case. With complete certainty we can say that out of the entire
sample of optically isolated galaxies there are no examples of
Arp-like bodily-distorted systems. 

There are a handful of isolated galaxies that are peculiar to some
lesser degree. But even this is to be expected without having to
invoke GNL galaxies; mergers will deplete the apparent population of
visible interactors while still leaving evidence of the collision in
the form of tidal debris or lingering asymmetries. For example, even a
cursory examination of the 2MASS near-infrared image of KIG 0022 (an
object that has large `tidal' arms in the optical) shows that it has a
double nucleus; presumably the result of a recent merger of two
(previously) visible galaxies.  Details of these samples and their
analysis will be given in a forthcoming paper (Madore. Petrillo \&
Nelson 2007).
\section{Conclusions}   

Using UV light as a tracer for star formation, FUV and NUV imaging
observations of nearby galaxies using the GALEX satellite show a
smooth and monotonic decline of star formation with total gas surface
density. No thresholding of star formation is visible in this sample,
at these projected surface densities.

The summary observations of peculiar galaxies viewed in the context of
$Lamda$-CDM simulations are as follows: (1) All cataloged ring
galaxies have plausible colliders that are optically visible. (2) All
peculiar galaxies (in the Arp Atlas) that are bodily deformed have
visible nearby companions that are plausibly responsible for the
interaction-induced deformities. (3) Virtually all isolated galaxies
are not peculiar, distorted or interacting to any noticeable degree.

The peculiar galaxy study leads to the following general conclusions:
(1) No ring galaxy is being produced from a head-on collision between
a spiral and pure dark-matter GNL galaxy (2) No bodily deformed galaxies
are the result of collisions and/or near encounters between optical
and pure dark-matter GNL galaxies. (3) Optically isolated galaxies
show no signs of bodily interactions with pure dark matter GNL
galaxies.


\vfill\eject
\acknowledgements 
We sincerely thank all of our colleagues on the GALEX Mission for their
support in enabling the science being conducted by NASA's Galaxy
Evolution Explorer. Major portions of this research would not have
been possible without the NASA/IPAC Extragalactic database (NED) which
is operated by the Jet Propulsion Laboratory, California Institute of
Technology, under contract with the National Aeronautics and Space
Administration. Significant support for this research was also
provided by the Observatories of the Carnegie Institution of
Washington.


\end{document}